\documentclass[a4paper,twoside]{article}

\usepackage{apalike}
\usepackage{amssymb}
\usepackage{amstext}
\usepackage{amsmath}
\usepackage{amsthm}
\usepackage{graphicx}
\usepackage{multicol}
\usepackage[small]{caption}

\usepackage{tikz}

\usepackage{SCITEPRESS}

\begin{document}


\title{RIOT OS Paves the Way for Implementation of High-Performance
       MAC Protocols}

\author{
\authorname{K\'evin Roussel, Ye-Qiong Song and Olivier Zendra}
\affiliation{LORIA/INRIA Nancy Grand-Est,\\
             Universit\'e de Lorraine,\\
             615, rue du Jardin Botanique,\\
             54600 Villers-L\`es-Nancy, France}
\email{\\
       \texttt{\{Kevin.Roussel,Ye-Qiong.Song,Olivier.Zendra\}@inria.fr}}
}


\abstract{Implementing new, high-performance MAC protocols requires
real-time features, to be able to synchronize correctly between different
unrelated devices. Such features are highly desirable for operating
wireless sensor networks (WSN) that are designed to be part of the Internet
of Things (IoT). Unfortunately, the operating systems commonly used in this
domain cannot provide such features. On the other hand, ``bare-metal''
development sacrifices portability, as well as the multitasking abilities
needed to develop the rich applications that are useful in the domain of
the Internet of Things.\\
We describe in this paper how we helped solving these issues by contributing 
to the development of a port of RIOT OS on the MSP430 microcontroller, an
architecture widely used in IoT-enabled motes. RIOT OS offers rich and
advanced real-time features, especially the simultaneous use of as many
hardware timers as the underlying platform (microcontroller) can offer.
We then demonstrate the effectiveness of these features by presenting
a new implementation, on RIOT OS, of S-CoSenS, an efficient MAC protocol
that uses very low processing power and energy.}

\onecolumn \maketitle \normalsize \vfill


\section{\uppercase{Introduction}}

When programming the small devices that constitutes the nodes of
the Internet of Things (IoT), one has to adapt to the limitations of
these devices.

Apart from their very limited processing power (especially compared to
the current personal computers, and even mobile devices like smartphones
and tablets), the main specificity of the devices is that they are operated
on small batteries (e.g.: AAA or button cells).

Thus, one of the main challenges with these motes is the need to reduce
as much as possible their energy consumption. We want their batteries to last
as long as possible, for economical but also practical reasons: it may be
difficult---even almost impossible---to change the batteries of some of these
motes, because of their locations (e.g.: on top of buildings, under roads,
etc.)

IoT motes are usually very compact devices: they are usually built around
a central integrated chip that contains the main processing unit and several
basic peripherals (such as timers, A/D and D/A converters, I/O
controllers\ldots) called microcontroller units or MCUs. Apart from the MCU,
a mote generally only contains some ``physical-world'' sensors and a radio
transceiver for networking. The main radio communication protocol currently
used in the IoT field is IEEE 802.15.4. Some MCUs do integrate a 802.15.4
transceiver on-chip.

Among the various components that constitute a mote, the most power-consuming
block is the radio transceiver. Consequently, to reduce the power consumption
of IoT motes, a first key point is to use the radio transceiver only when
needed, keeping it powered-off as much as possible. The software element
responsible to control the radio transceiver in an adequate manner is
the \emph{MAC~/ RDC (Media Access Control \& Radio Duty Cycle)}
layer of the network stack.

A efficient power-saving strategy for IoT motes thus relies on finding the
better trade-off between minimizing the radio duty cycle while keeping
networking efficiency at the highest possible level. This is achieved
by developing new, ``intelligent'' MAC~/ RDC protocols.

To implement new, high-performance MAC~/ RDC protocols, one needs to be
able to react to events with good reactivity (lowest latency possible) and
flexibility. These protocols rely on precise timing to ensure efficient
synchronization between the different motes and other radio-networked
devices of a \emph{Personal Area Network (PAN)}, thus allowing
to turn on the radio transceivers \emph{only} when needed.

At the system level, being able to follow such accurate timings means having
very efficient interruption management, and the extensive use of hardware
timers, that are the most precise timing source available.

The second most power-consuming element in a mote, after the radio
transceiver, is the MCU itself: every current MCU offers ``low-power modes'',
that consist in disabling the various hardware blocks, beginning with the CPU
core. The main way to minimize energy consumption with a MCU is thus
to disable its features as much as possible, only using them when needed:
that effectively means putting the whole MCU to sleep as much as possible.

Like for the radio transceiver, using the MCU efficiently while keeping
the system efficient and reactive means optimal use of interruptions,
and hardware timers for synchronization.

Thus, in both cases, we need to optimally use interruptions as well as
hardware timers. Being able to use them both efficiently without too much
hassle implies the use of a specialized operating system (OS), especially
to easily benefit from multitasking abilities. That is what we will
discuss in this paper.


\section{\uppercase{Previous work and problem statement}}

Specialized OSes for the resource-constrained devices that constitute
wireless sensor networks have been designed, published, and made available
for quite a long time.

\subsection{TinyOS}

The first widely used system in this domain was \emph{TinyOS} \cite{TinyOS}.
It is an open-source OS, whose first stable release (1.0) was published in
september 2002. It is very lightweight, and as such well adapted to limited
devices like WSN motes.  It has brought many advances in this domain, like
the ability to use Internet Protocol (IP) and routing (RPL) on 802.15.4
networks, including the latest IPv6 version, and to simulate networks
of TinyOS motes via TOSSIM \cite{TOSSIM}.

Its main drawback is that one needs to learn a specific language---named
nesC---to be able to efficiently work within it. This language is quite
different from standard C and other common imperative programming languages,
and as such can be difficult to master.

The presence of that specific language is no coincidence: TinyOS is built
on its own specific paradigms: it has an unique stack, from which the
different components of the OS are called as statically linked callbacks.
This makes the programming of applications complex, especially for
decomposing into various ``tasks''. The multitasking part is also
quite limited: tasks are run in a fixed, queue-like order. Finally,
TinyOS requires a custom GNU-based toolchain to be built.

All of these limitations, plus a relatively slow development pace (last
stable version dates back to august 2012) have harmed its adoption,
and it is not the mainly used OS of the domain anymore.

\subsection{Contiki}

The current reference OS in the domain of WSN and IoT is \emph{Contiki}
\cite{ContikiOS}. It's also an open-source OS, which was first released
in 2002. It is also at the origin of many assets: we can mention, among
others, the uIP Embedded TCP/IP Stack \cite{uip}, that has been extended
to uIPv6, the low-power Rime network stack \cite{Rime}, or the Cooja advanced
network simulator \cite{Cooja}.

While a bit more resource-demanding than TinyOS, Contiki is also very
lightweight and well adapted to motes. Its greatest advantage over TinyOS
is that it is based on standard, well-known OS paradigms, and coded
in standard C language, which makes it relatively easy to learn and program.
It offers an event-based kernel, implemented using cooperative multithreading,
and a complete network stack. All of these features and advantages have made
Contiki widespread, making it the reference OS when it comes to WSN.

Contiki developers also have made advances in the MAC/RDC domain: many
of them have been implemented as part of the Contiki network stack, and
a specifically developed, ContikiMAC, has been published in 2011
\cite{ContikiMAC} and implemented into Contiki as the default
RDC protocol (designed to be used with standard CSMA/CA as MAC layer).

However, Contiki's extremely compact footprint and high optimization comes
at the cost of some limitations that prevented us from using it as our
software platform.

Contiki OS is indeed not a real-time OS: the processing of ``events''---using
Contiki's terminology---is made by using the kernel's scheduler, which is
based on cooperative multitasking. This scheduler only triggers at a specific,
pre-determined rate; on the platforms we're interested in, this rate is
fixed to 128~Hz: this corresponds to a time skew of up to 8~milliseconds
(8000~microseconds) to process an event, interruption management being
one of the possible events. Such a large granularity is clearly
a huge problem when implementing high-performance MAC/RDC protocols,
knowing that the transmission of a full-length 802.15.4 packet takes
bout 4~milliseconds (4000~microseconds), a time granularity of
320~microseconds is needed, corresponding to one backoff period (BP).

To address this problem, Contiki provides a real-time feature,
\texttt{rtimer}, which allows to bypass the kernel scheduler and use
a hardware timer to trigger execution of user-defined functions. However,
it has very severe limitations:

\begin{itemize}

\item only one instance of \texttt{rtimer} is available, thus only one
real-time event can be scheduled or executed at any time; this limitation
forbids development of advanced real-time software---like high-performance
MAC~/ RDC protocols---or at least makes it very hard;

\item moreover, it is unsafe to execute from \texttt{rtimer}, even
indirectly, most of the Contiki basic functions (i.e.: kernel, network
stack, etc.), because these functions are not designed to handle pre-emption.
Contiki is indeed based on cooperative multithreading, whereas the
\texttt{rtimer} mechanism seems like a ``independent feature'', coming
with its own paradigm.
Only a precise set of functions known as ``interrupt-safe'' (like
\texttt{process\_poll()}) can be safely invoked from \texttt{rtimer},
using other parts of Contiki's meaning almost certainly crash or
unpredictable behaviour. This restriction practically makes it very
difficult to write Contiki extensions (like network stack layer drivers)
using \texttt{rtimer}.


\end{itemize}

Also note that this cooperative scheduler is designed to manage a specific
kind of tasks: the \emph{protothreads}. This solution allows to manage
different threads of execution, without needing each of them to have
its own separate stack \cite{Protothreads}. The great advantage of
this mechanism is the ability to use an unique stack, thus greatly
reducing the needed amount of RAM for the system. The trade-off is
that one must be careful when using certain C constructs (i.e.:
it is impossible to use the \texttt{switch} statement in
some parts of programs that use protothreads).

For all these reasons, we were unable to use Contiki OS to develop and
implement our high-performance MAC/RDC protocols. We definitely needed
an OS with efficient real-time features and event handling mechanism.

\subsection{Other options}

There are other, less used OSes designed for the WSN/IoT domain, but none
of them fulfilled our requirements, for the following reasons:
\begin{description}

\item[SOS] \cite{SOS} This system's development has been cancelled since november
           2008; its authors explicitly recommend on their website to
           ``consider one of the more actively supported alternatives''.

\item[Lorien] \cite{LorienOS} While its component-oriented approach is
              interesting, this system seems does not seem very widespread. It is currently available for only 
              one hardware platform (TelosB/SkyMote) which seriously
              limits the portability we can expect from using an OS. 
              Moreover, its development seems to have slowed down quite
              a bit, since the latest available Lorien release was published
              in july 2011, while the latest commit in the project's
              SourceForge repository (r46) dates back to january 2013.

\item[Mantis] \cite{MantisOS} While this project claims to be Open Source,
              the project has made, on its SourceForge web site, no public
              release, and the access to the source repository 
              (\texttt{http://mantis.cs.colorado.edu/viewcvs/}) seems
              to stall. Moreover, reading the project's main web page
              shows us that the last posted news item mentions a first beta
              to be released in 2007. The last publications about
              Mantis OS also seems to be in 2007. All of these elements 
              tend to indicate that this project is abandoned\ldots

\item[LiteOS] \cite{LiteOS} This system offers very interesting features,
              especially the ability to update the nodes firmwares over the wireless,
              as well as the built-in hierarchical file system. Unfortunately,
              it is currently only available on IRIS/MicaZ platforms,
              and requires AVR Studio for programming (which imposes
              Microsoft Windows as a development platform). This
              greatly hinders portability, since LiteOS is clearly strongly
              tied to the AVR microcontroller architecture.

\item[MansOS] \cite{MansOS} This system is very recent and offers many
              interesting features, like optional preemptive multitasking,
              a network stack, runtime reprogramming, and a scripting
              language. It is available on two MCU architectures: AVR and
              MSP430 (but not ARM). However, none of the real-time features
              we wanted seems to be available: e.g. only software timers with
              a 1~millisecond resolution are available.

\end{description}
In any case, none of the alternative OSes cited hereabove offer the real-time
features we were looking for.

\bigskip

On the other hand, ``bare-metal'' programming is also unacceptable for us:
it would mean sacrificing portability and multitasking; and we would also
need to redevelop many tools and APIs to make application programming
even remotely practical enough for third-party developers who would
want to use our protocols.

\bigskip

We also envisioned to use an established real-time OS (RTOS) as a base
for our works. The current reference when it comes to open-source RTOS is
\emph{FreeRTOS} (\texttt{http://www.freertos.org/}). It is a robust, mature
and widely used OS. Its codebase consists in clean and well-documented
standard C language. However, it offers only core features, and doesn't
provide any network subsystem at all. Redeveloping a whole network stack
from scratch would have been too time-consuming.
(Network extensions exist for FreeRTOS, but they are either immature,
or very limited, or proprietary and commercial software; and most of them
are tied to a peculiar piece of hardware, thus ruining
the portability advantage offered by the OS.)

\subsection{Summary: Wanted Features}

To summarize the issue, what we required is an OS that:
\begin{itemize}
\item is adapted to the limitations of the deeply-embedded MCUs that
      constitute the core of WSN/IoT motes;
\item provides real-time features powerful enough to support the
      development of advanced, high-performance MAC~/ RDC protocols;
\item includes a network stack (even a basic one) adapted to wireless
      communication on 802.15.4 radio medium.
\end{itemize}
However, none of the established OSes commonly used either in the IoT domain
(TinyOS, Contiki) nor in the larger spectrum of RTOS (FreeRTOS)
could match our needs.


\section{\uppercase{The RIOT Operating System}}

Consequently, we focused our interest on \emph{RIOT OS} \cite{RIOT}.

This new system---first released in 2013---is also open-source and
specialized in the domain of low-power, embedded wireless sensors.
It offers many interesting features, that we will now describe.

It provides the basic benefits of an OS: portability (it has been ported
to many devices powered by ARM, MSP430, and---more recently---AVR
microcontrollers) and a comprehensive set of features, including
a network stack.

Moreover, it offers key features that are otherwise yet unknown in
the WSN/IoT domain:

\begin{itemize}

\item an efficient, interrupt-driven, tickless \emph{micro-kernel};

\item that kernel includes a priority-aware task scheduler, providing
      \emph{pre-emptive multitasking};

\item a highly efficient use of \emph{hardware timers}: all of them can be
      used concurrently (especially since the kernel is tickless), offering
      the ability to schedule actions with high granularity; on low-end
      devices, based on MSP430 architecture, events can be scheduled
      with a resolution of 32~microseconds;

\item RIOT is entirely written in \emph{standard C language}; but unlike
      Contiki, there are no restrictions on usable constructs (i.e.: like
      those introduced by the protothreads mechanism);

\item a clean and \emph{modular design}, that makes development with and
      \emph{into} the system itself easier and more productive.

\end{itemize}

The first three features listed hereabove make RIOT a full-fledged
\emph{real-time} operating system.

We also believe that the tickless kernel and the optimal use of hardware
timers should make RIOT OS a very suited software platform to optimize energy
consumption on battery-powered, MCU-based devices.

A drawback of RIOT, compared to TinyOS or Contiki, is its higher memory
footprint: the full network stack (from PHY driver up to RPL routing with
\mbox{6LoWPAN} and MAC~/ RDC layers) cannot be compiled for Sky/TelosB
because of overflowing memory space. Right now, constrained devices like
MSP430-based motes are limited to the role of what the 802.15.4 standard
calls \emph{Reduced Function Devices (RFD)}, the role of \emph{Full
Function Devices (FFD)} being reserved to more powerful motes (i.e.:
based on ARM microcontrollers).

However, we also note that, thanks to its modular architecture, the RIOT
kernel, compiled with only PHY and MAC~/ RDC layers, is actually lightweight
and consumes little memory. We consequently believe that the current
situation will improve with the maturation of higher layers of RIOT network
stack, and that in the future more constrained devices could also be used
as FFD with RIOT OS.

\medskip

When we began to work with RIOT, it also had two other issues: the MSP430
versions were not stable enough to make real use of the platform; and
beyond basic CSMA/CA, no work related to the MAC~/ RDC layer had been
done on that system. This is where our contributions fit in.


\section{\uppercase{Our contributions}}

For our work, we use---as our main hardware platform---IoT motes built
around MSP430 microcontrollers.

MSP430 is a microcontroller (MCU) architecture from Texas Instruments,
offering very low-power consumption, cheap price, and good performance thanks
to a custom 16-bit RISC design. This architecture is very common in IoT motes.
It is also very well supported, especially by the Cooja simulator
\cite{Cooja}, which makes simulations of network scenarios---especially
with many devices---much easier to design and test.

RIOT OS has historically been developed first on legacy ARM devices
(ARM7TDMI-based MCUs), then ported on more recent microcontrollers
(ARM Cortex-M) and other architectures (MSP430 then AVR). However,
the MSP430 port was, before we improved it, still not as ``polished''
as ARM code and thus prone to crash.

Our contribution can be summarized in the following points:

\begin{enumerate}

\item analysis of current OSes (TinyOS, Contiki, etc.) limitations,
      and why they are incompatible with development of real-time
      extensions like advanced MAC~/ RDC protocols;

\item add debugging features to the RIOT OS kernel, more precisely
      a mechanism to handle fatal errors: crashed systems can be
      ``frozen'' to facilitate debugging during development; or,
      in production, can be made to reboot immediately, thus reducing
      unavailability of a RIOT-running device to a minimum;

\item port RIOT OS to a production-ready, MSP430-based device:
      the Zolertia Z1 mote (already supoorted by Contiki,
      and used in real-world scenarios running that OS);

\item debug the MSP430-specific portion of RIOT OS---more specifically:
      the hardware abstraction layer (HAL) of the task scheduler---making
      RIOT OS robust and production-ready on MSP430-based devices.\\
      Note that all of  these contributions have been reviewed by RIOT's
      development team and integrated into the ``master'' branch of RIOT OS'
      Github repository (i.e.: they are now part of the standard code base of
      the system).

\item running on MSP430-based devices also allows RIOT OS applications
      to be simulated with the Cooja simulator; this greatly improves
      speed and ease of development.

\item thanks to these achievements, we now have a robust and full-featured
      software platform offering all the features needed to develop
      high-performance MAC/RDC protocols---such as all of the time-slotted
      protocols.

\end{enumerate}

As a proof of concept of this last statement, we have implemented one
of our own designs, and obtained very promising results, shown in
the next section.


\section{\uppercase{Use Case: implementing the S-CoSenS RDC protocol}}

\subsection{The S-CoSenS Protocol}

The first protocol we wanted to implement is S-CoSenS \cite{TheseBNefzi},
which is designed to work on top of the IEEE 802.15.4 physical and MAC
(i.e.: CSMA/CA) layers.

It is an evolution of the already published CoSenS protocol \cite{CosensConf}:
it adds to the latter a sleeping period for energy saving.
Thus, the basic principle of S-CoSenS is to delay the forwarding (routing)
of received packets, by dividing the radio duty cycle in three periods:
a sleeping period (SP), a waiting period (WP) where the radio medium
is listened by routers for collecting incoming 802.15.4 packets, and
finally a burst transmission period (TP) for emitting adequately
the packets enqueued during WP.

The main advantage of S-CoSenS is its ability to adapt dynamically to the
wireless network throughput at runtime, by calculating for each radio duty
cycle the length of SP and WP, according to the number of relayed
packets during previous cycles. Note that the set of the SP and the WP
of a same cycle is named \emph{subframe}; it is the part of a S-CoSenS
cycle whose length is computed and known \textit{a priori}; on the contrary,
TP duration is always unknown up to its very beginning, because it depends
on the amount of data successfully received during the WP that precedes it.

The computation of WP duration follows a ``sliding average'' algorithm,
where WP duration for each duty cycle is computed from the average
of previous cycles as:
\begin{eqnarray*}
&&
\overline{\mathrm{WP}_{n}} = \alpha \cdot \overline{\mathrm{WP}_{n-1}}
                + (1 - \alpha) \cdot \mathrm{WP}_{n-1}
\\ &&
\mathrm{WP}_{n} = \max ( \mathrm{WP}_{min},
                  \min ( \overline{\mathrm{WP}_{n}}, \mathrm{WP}_{max} ) )
\end{eqnarray*}
where $\overline{\mathrm{WP}_{n}}$ and $\overline{\mathrm{WP}_{n-1}}$
are respectively the average WP length at $n^{\mathrm{th}}$ and
$(n-1)^{\mathrm{th}}$ cycle, while $\mathrm{WP}_{n}$ and $\mathrm{WP}_{n-1}$
are the actual length of respectively the $n^{\mathrm{th}}$ and
$(n-1)^{\mathrm{th}}$ cycles; $\alpha$ is a parameter between 0 and 1
representing the relative weight of the history in the computation,
and $\mathrm{WP}_{min}$ and $\mathrm{WP}_{max}$ are high and low limits
imposed by the programmer to the WP duration.

The length of the whole subframe being a parameter given at compilation time,
SP duration is simply computed by subtracting the calculated duration of WP
from the subframe duration for every cycle.

The local synchronization between a S-CoSenS router and its leaf nodes
is done thanks to a beacon packet, that is broadcasted by the router at
the beginning of each cycle. This beacon contains the duration
(in microseconds) of the SP and WP for the currently
beginning cycle.

\begin{figure}[!ht]
\centering
\begin{tikzpicture}[>=latex]
\fill[black] (0cm, -0.25cm) rectangle +(0.2cm, 0.5cm);
\draw[->,thick] (0.1cm, 0.25cm) -- +(0, 0.5cm);
\draw (0.1cm, 1.3cm) node {Beacon};
\draw[anchor=west] (-0.6cm, 0.9cm) node {(broadcasted)};
\draw[thick] (0cm, -0.25cm) -- +(0, 0.5cm);
\foreach \x in {1,2,3,4,5,6}
{
  \fill[lightgray] (0.2cm + \x * 0.25cm, -0.25cm) rectangle +(0.05cm, 0.5cm);
}
\draw (1.1cm, 0) node {\textbf{SP}};
\draw[thick] (2cm, -0.25cm) -- +(0, 0.5cm);
\fill[lightgray] (2cm, -0.25cm) rectangle +(2cm, 0.5cm);
\draw (3cm, 0) node {\textbf{WP}};
\draw[thick] (4cm, -0.25cm) -- +(0, 0.5cm);
\fill[lightgray] (4cm, -0.25cm) rectangle +(2cm, 0.5cm);
\draw (5cm, 0) node {\textbf{TP}};
\draw[thick] (6cm, -0.25cm) -- +(0, 0.5cm);
\draw[->] (-0.5cm,  0.25cm) -- +(7cm, 0);
\draw[->] (-0.5cm, -0.25cm) -- +(7cm, 0);
\draw[->,thick] (2.5cm, 0.75cm) -- +(0, -0.5cm);
\draw (2.5cm, 1cm) node {P1};
\draw[->,thick] (3cm, 0.75cm) -- +(0, -0.5cm);
\draw (3cm, 1cm) node {P2};
\draw[->,thick] (3.5cm, 0.75cm) -- +(0, -0.5cm);
\draw (3.5cm, 1cm) node {P3};
\draw[->,thick] (4.5cm, 0.25cm) -- +(0, 0.5cm);
\draw (4.5cm, 1cm) node {P1};
\draw[->,thick] (5cm, 0.25cm) -- +(0, 0.5cm);
\draw (5cm, 1cm) node {P2};
\draw[->,thick] (5.5cm, 0.25cm) -- +(0, 0.5cm);
\draw (5.5cm, 1cm) node {P3};
\draw (0cm, -0.5cm)  .. controls +(0, -0.25cm) .. +(1cm, -0.25cm);
\draw (1cm, -0.75cm) .. controls +(1cm, 0)     .. +(1cm, -0.25cm);
\draw (2cm, -1cm)    .. controls +(0, 0.25cm)  .. +(1cm, 0.25cm);
\draw (3cm, -0.75cm) .. controls +(1cm, 0)     .. +(1cm, 0.25cm);
\draw (2cm, -1.25cm) node {\textbf{Subframe}};
\draw (0cm, -1.5cm)    .. controls +(0, -0.25cm) .. +(1.5cm, -0.25cm);
\draw (1.5cm, -1.75cm) .. controls +(1.5cm, 0)   .. +(1.5cm, -0.25cm);
\draw (3cm, -2cm)      .. controls +(0, 0.25cm)  .. +(1.5cm, 0.25cm);
\draw (4.5cm, -1.75cm) .. controls +(1.5cm, 0)   .. +(1.5cm, 0.25cm);
\end{tikzpicture}
\caption{A typical S-CoSenS router cycle.\\
         The gray strips in the SP represents the short wake-up-and-listen
         periods used for inter-router communication.}
\label{FigSCosensDutyCycle}
\end{figure}
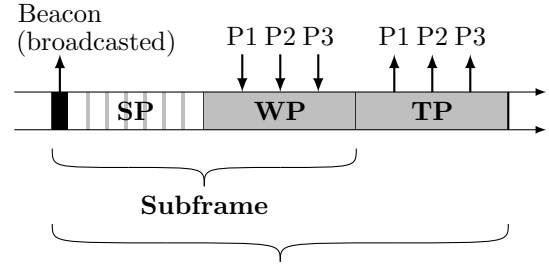

The whole S-CoSenS cycle workflow for a router is summarized in figure
\ref{FigSCosensDutyCycle} hereafter.

\begin{figure}[!h]
\centering
\begin{tikzpicture}[>=latex]
\draw (-0.5cm, 0) node {\large \textit{R}};
\draw[thick] (1cm, -0.25cm) -- +(0, 0.5cm);
\draw (2cm, 0) node {\textbf{SP}};
\draw[thick] (3cm, -0.25cm) -- +(0, 0.5cm);
\fill[lightgray] (3cm, -0.25cm) rectangle +(2cm, 0.5cm);
\draw (4cm, 0) node {\textbf{WP}};
\draw[thick] (5cm, -0.25cm) -- +(0, 0.5cm);
\fill[lightgray] (5cm, -0.25cm) rectangle +(0.5cm, 0.5cm);
\draw (5.25cm, -0.5cm) node {\textbf{TP}};
\draw[thick] (5.5cm, -0.25cm) -- +(0, 0.5cm);
\draw[->] (-0.5cm,  0.25cm) -- +(6.5cm, 0);
\draw[->] (-0.5cm, -0.25cm) -- +(6.5cm, 0);
\draw (-0.5cm, -1.5cm) node {\large \textit{LN}};
\fill[gray] (0cm, -1.25cm) rectangle +(1.3cm, -0.5cm);
\fill[gray] (2.9cm, -1.25cm) rectangle +(0.5cm, -0.5cm);
\fill[black] (1cm, -0.25cm) rectangle +(0.2cm, 0.5cm);
\draw[->,thick] (1.1cm, 0.25cm) -- +(0, -1.5cm);
\draw[anchor=east] (1cm, -0.75cm) node {Beacon};
\fill[black] (1cm, -1.25cm) rectangle +(0.2cm, -0.5cm);
\draw[->,very thick] (0cm, -2.5cm) -- +(0, 0.75cm);
\draw[anchor=west] (0cm, -2.5cm)
     node {\footnotesize \textbf{packet arrival}};
\fill[black] (3.1cm, -1.25cm) rectangle +(0.2cm, -0.5cm);
\draw[->,thick] (3.2cm, -1.25cm) -- +(0, 1cm);
\draw[anchor=west] (3.2cm, -0.75cm) node {P1};
\fill[black] (3.1cm, -0.25cm) rectangle +(0.2cm, 0.5cm);
\fill[black] (5.1cm, -0.25cm) rectangle +(0.2cm, 0.5cm);
\draw[->,thick] (5.2cm, 0.25cm) -- +(0, 0.5cm);
\draw (5.2cm, 1cm) node {P1};
\draw[->] (-0.5cm, -1.25cm) -- +(6.5cm, 0);
\draw[->] (-0.5cm, -1.75cm) -- +(6.5cm, 0);
\end{tikzpicture}
\caption{A typical transmission of a data packet with the S-CoSenS protocol
         between a leaf node and a router.}
\label{FigSCoSenSPktTx}
\end{figure}
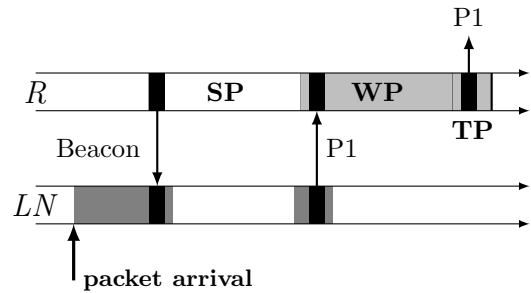

An interesting property of S-CoSenS is that leaf (i.e.: non-router) nodes
always have their radio transceiver offline, except when they have packets
to send. When a data packet is generated on a leaf node, the latter wakes up
its radio transceiver, listens and waits to the first beacon emitted by
an S-CoSenS router, then sends its packet using CSMA/CA at the beginning
of the WP described in the beacon it received. A leaf node will put its
transceiver offline during the delay between the beacon and that WP
(that is: the SP of the router that emitted the received beacon), and
will go back to sleep mode once its packet is transmitted.
All of this procedure is shown in figure \ref{FigSCoSenSPktTx}.

We thus need to synchronize with enough accuracy different devices (that
can be based on different hardware platforms) on cycles whose periods
are dynamically calculated at runtime, with resolution that needs to be
in the sub-millisecond range. This is where RIOT OS advanced real-time
features really shine, while the other comparable OSes are
for that purpose definitely lacking.

\subsection{Simulations and Synchronization Accuracy}

\begin{figure}[!h]
\centering
\begin{tikzpicture}[>=latex]
\draw (0, 1cm) circle (0.25cm); \draw (0, 1cm) node {S};
\draw[->,thick] (0, 0.25cm) -- (0, 0.75cm);
\draw (0, 0) circle (0.25cm); \draw (0, 0) node {R};
\foreach \x in {6,7,8,9,10}
{
  \fill[white] (\x * 1cm - 8cm, -1.75cm) circle (0.25cm);
  \draw (\x * 1cm - 8cm, -1.75cm) circle (0.25cm);
  \draw (\x * 1cm - 8cm, -1.75cm) node {\x};
  \draw[->,thick] (\x * 1cm - 8cm, -1.5cm)
                  -- (\x * 0.02cm - 0.16cm, -0.25cm);
}
\foreach \x in {1,2,3,4,5}
{
  \fill[white] (\x * 1cm - 3cm, -1cm) circle (0.25cm);
  \draw (\x * 1cm - 3cm, -1cm) circle (0.25cm);
  \draw (\x * 1cm - 3cm, -1cm) node {\x};
  \draw[->,thick] (\x * 1cm - 3cm, -0.75cm)
                  -- (\x * 0.05cm - 0.15cm, -0.25cm);
}
\end{tikzpicture}
\caption{Functional schema of our virtual test PAN.}
\label{FigPANtest}
\end{figure}
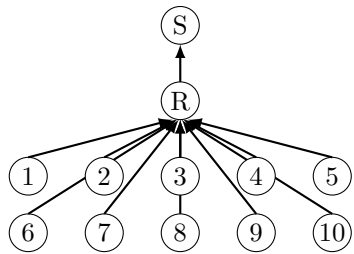

We have implemented S-CoSenS under RIOT, and made first tests by performing
simulations---with Cooja---of a 802.15.4 PAN (Personal Area Network)
constituted of a router, and ten motes acting as ``leaf nodes''.
The ten nodes regularly send data packets to the router, that retransmits
these data packets to a nearby ``sink'' device. Both the router and the ten
nodes use exclusively the S-CoSenS RDC/MAC protocol. This is summarized
in figure \ref{FigPANtest}.

\begin{figure*}[ptb]
\centering
\includegraphics[width=15.75cm]{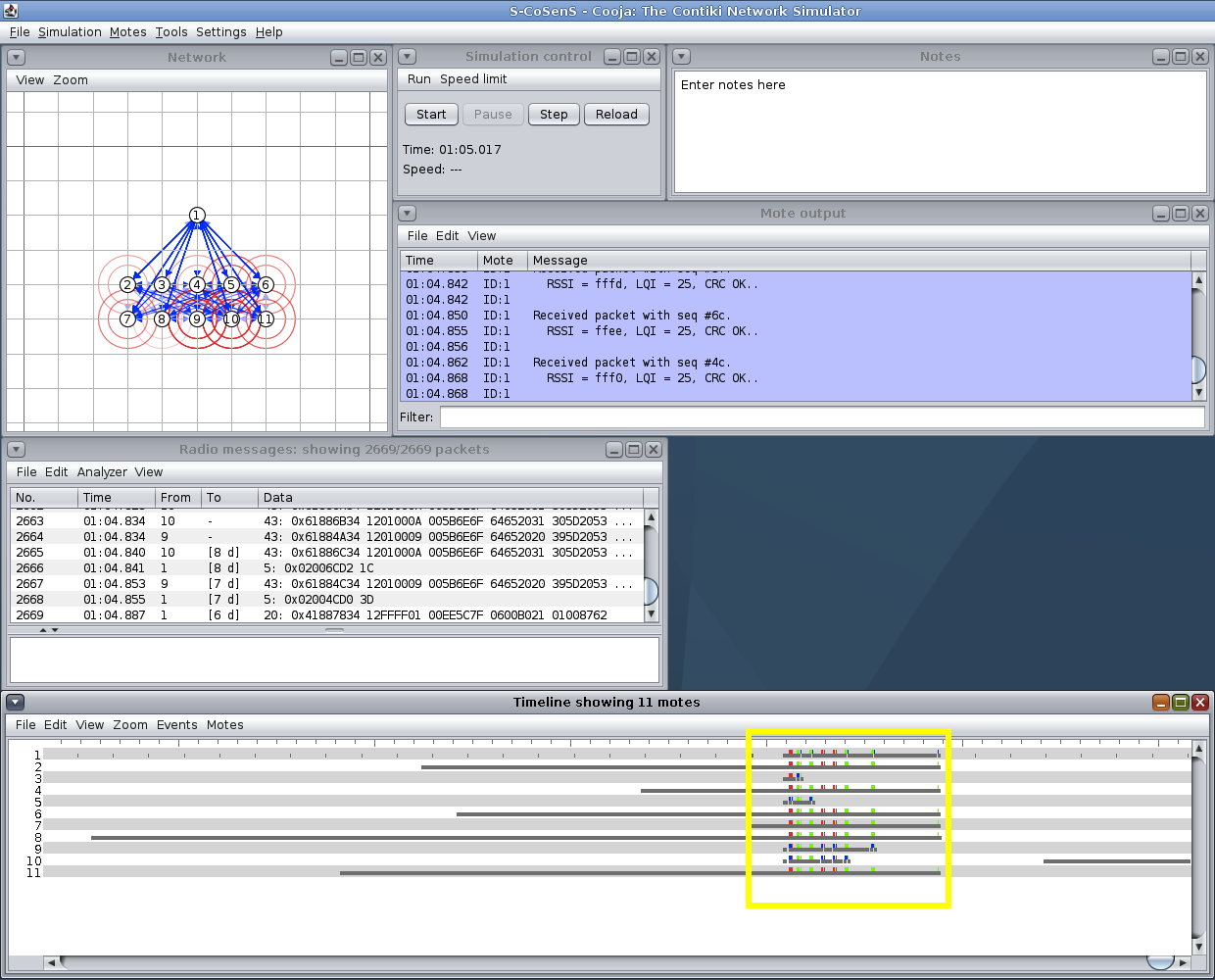}
\caption{Screenshot of our test simulation in Cooja. 
(Despite the window title mentioning Contiki, the simulated application
 is indeed running on RIOT OS.)}
\label{Screenshot}
\end{figure*}

\begin{figure*}[pbt]
\centering
\includegraphics{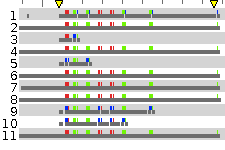}
\caption{Zoom on the central part of the timeline of our simulation.}
\label{ZoomTimeline}
\end{figure*}

Our first tests clearly show an excellent synchronization between the
leaf nodes and the router, thanks to the time resolution offered by RIOT OS
event management system (especially the availability of many hardware
timers for direct use). This can be seen in the screenshot of our
simulation in Cooja, shown in figure \ref{Screenshot}. For readability,
the central portion of the timeline window of that screenshot (delimited
by a thick yellow rectangle) is zoomed on in figure \ref{ZoomTimeline}.

On figure \ref{ZoomTimeline}, the numbers on the left side are motes'
numerical IDs: the router has ID number \textsf{1}, while the leaf nodes
have IDs \textsf{2} to \textsf{11}. Grey bars represent radio transceiver
being online for a given mote; blue bars represent packet emission, and green
bars correct packet reception, while red bars represent collision (when
two or more devices emit data concurrently) and thus reception of
undecipherable radio signals.

Figure \ref{ZoomTimeline} represents a short amount of time (around
100~milliseconds), representing the end of a duty cycle of the router:
the first 20~milliseconds are the end of SP, and 80 remaining milliseconds
the WP, then the beginning of a new duty cycle (the TP has been disabled
in our simulation). 

In our example, four nodes have data to transmit to the router: the motes
number \textsf{3}, \textsf{5}, \textsf{9}, and \textsf{10}; the other nodes
(\textsf{2}, \textsf{4}, \textsf{6}, \textsf{7}, \textsf{8}, and \textsf{11})
are preparing to transmit a packet in the next duty cycle.

At the instant marked by the first yellow arrow (in the top left of figure
\ref{ZoomTimeline}), the SP ends and the router activates its radio
transceiver to enter WP. Note how the four nodes that are to send packets
(\textsf{3}, \textsf{5}, \textsf{9}, and \textsf{10}) do also activate their
radio transceivers \emph{precisely} at the same instant: this is thanks to
RIOT OS precise real-time mechanism (based on hardware timers), that allows
to the different nodes to precisely synchronize on the timing values
transmitted in the previous beacon packet. Thanks also to that mechanism,
the nodes are able to keep both their radio transceiver \emph{and} their
MCU in low-power mode, since RIOT OS kernel is interrupt-driven.

During the waiting period, we also see that several collisions occur; they
are resolved by the S-CoSenS protocol by forcing motes to wait a random
duration before re-emitting a packet in case of conflict. In our example,
our four motes can finally transmit their packet to the router in that
order: \textsf{3} (after a first collision), \textsf{5}, \textsf{10} (after
two other collisions), and finally \textsf{9}. Note that every time the
router (device number \textsf{1}) successfully receives a packet, an
acknowledgement is sent back to emitter: see the very thin blue bars that
follow each green bar on the first line.

Finally, at the instant marked by the second yellow arrow (in the top right
of figure \ref{ZoomTimeline}), WP ends and a new duty cycle begins.
Consequently, the router broadcasts a beacon packet containing PAN timing and
synchronization data to all of the ten nodes. We can see that all of the
six nodes waiting to transmit (\textsf{2}, \textsf{4}, \textsf{6}, \textsf{7},
\textsf{8}, and \textsf{11}) go idle after receiving this beacon (beacon
packets are broadcasted and thus not to be acknowledged): they go
into low-power mode (both at radio transceiver and MCU level), and will
take advantage of RIOT real-time features to wake up precisely when
the router goes back into WP mode and is ready to receive their
packets.

\subsection{Performance Evaluation: Preliminary Results}

We will now present the first, preliminary results we obtained through the
simulations we described hereabove.

Important: note that \emph{we evaluate here the implementations}, and not the
intrinsic advantages or weaknesses of the protocols themselves.

We have first focused on QoS results, by computing Packet Reception Rates
and end-to-end delays between the various leaf nodes and the sink of the test
PAN presented earlier in figure \ref{FigPANtest}, to evaluate the quality
of the transmissions allowed by using both of the protocols.

For these first tests, we used default parameters for both RDC protocols
(ContikiMAC and S-CoSenS), only pushing the CSMA/CA MAC layer of Contiki
to make up to 8 attempts for transmitting a same packet, so as to put it
on par with our implementation on RIOT OS. We have otherwise not yet
tried to tweak the various parameters offered by both the RDC protocols
to optimize results. This will be the subject of our next experiences.

\subsubsection{Packet Reception Rates (PRR)}

\begin{figure}
  \centering
  \includegraphics[width=7.5cm]{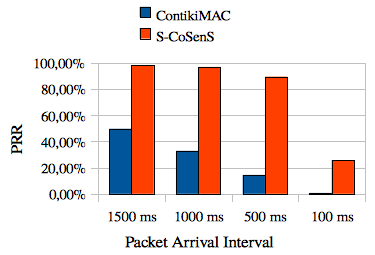}
  \caption{PRR results for both ContikiMAC and S-CoSenS RDC protocols,
           using default values for parameters.}
  \label{FigPRRresults}
\end{figure}

The result obtained for PRR using both protocols are shown in figure
\ref{FigPRRresults} as well as table \ref{TblPRRresults}.

\begin{table}
\centering
\begin{tabular}{|r|r|r|}
\hline
 PAI \textbackslash\ Protocol & ContikiMAC & S-CoSenS \\
\hline
 1500 ms & 49.70\% & 98.10\% \\
 1000 ms & 32.82\% & 96.90\% \\
  500 ms & 14.44\% & 89.44\% \\
  100 ms &  0.64\% & 25.80\% \\
\hline
\end{tabular}
\caption{PRR results for both ContikiMAC and S-CoSenS RDC protocols,
         using default values for parameters.}
\label{TblPRRresults}
\end{table}

The advantage of S-CoSenS as shown on the figure is clear and significant
whatever the packet arrival interval constated. Excepted for the ``extreme''
scenario corresponding to an over-saturation of the radio channel, S-CoSenS
achieve an excellent PRR ($\gtrapprox 90\%$), while ContikiMAC's PRR
is always $\lessapprox 50\%$.

\subsubsection{End-To-End Transmission Delays}

\begin{figure}
  \centering
  \includegraphics[width=7.5cm]{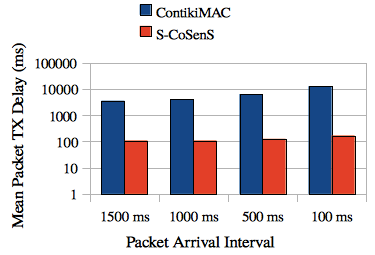}
  \caption{End-to-end delays results for both ContikiMAC and S-CoSenS RDC
           protocols, using default values for parameters; note that
           vertical axis is drawn with logarithmic scale.}
  \label{FigDelaysResults}
\end{figure}

The result obtained for PRR using both protocols are shown in figure
\ref{FigDelaysResults} and table \ref{TblDelaysResults}.

\begin{table}
\centering
\begin{tabular}{|r|r|r|}
\hline
 PAI \textbackslash\ Protocol & ContikiMAC & S-CoSenS \\
\hline
 1500 ms &  3579 ms & 108 ms \\
 1000 ms &  4093 ms & 108 ms \\
  500 ms &  6452 ms & 126 ms \\
  100 ms & 12913 ms & 168 ms \\
\hline
\end{tabular}
\caption{End-to-end delays results for both ContikiMAC and S-CoSenS RDC
         protocols, using default values for parameters.}
\label{TblDelaysResults}
\end{table}

S-CoSenS has here also clearly the upper hand, so much that we had to use
logarithmic scale for the vertical axis to keep figure \ref{FigDelaysResults}
easily readable. The advantage of S-CoSenS is valid whatever the packet
arrival interval, our protocol being able to keep delay below an acceptable
limit (in the magnitude of hundreds of milliseconds), while ContikiMAC
delays rocket up to tens of seconds when network load increases.

\subsubsection{Summary: QoS Considerations}

While these are only preliminary results, it seems that being able to
leverage real-time features is clearly a significant advantage when designing
and implementing MAC/RDC protocols, at least when it comes to QoS results.


\section{\uppercase{Future Works and Conclusion}}

We plan, in a near future:

\begin{itemize}

\item to bring new contributions to the RIOT project: we are especially
      interested in the portability that the RIOT solution offers us;
      this OS is indeed actively ported on many devices based on powerful
      microcrontrollers based on ARM Cortex-M architecture (especially
      Cortex-M3 and Cortex-M4), and we intend to help in this porting
      effort, especially on high-end IoT motes we seek to use in our
      works (e.g.: as advanced FFD nodes with full network stack,
      or routers);

\item to use the power of this OS to further advance our work on MAC/RDC
      protocols; more precisely, we are implementing other innovative
      MAC/RDC protocols---such as iQueue-MAC \cite{iQueueMAC}---under RIOT,
      taking advantage of its high-resolution real-time features to obtain
      excellent performance, optimal energy consumption, and out-of-the-box
      portability.

\end{itemize}

RIOT is a powerful real-time operating system, adapted to the limitations
of deeply embedded hardware microcontrollers, while offering state-of-the-art
techniques (preemptive multitasking, tickless scheduler, optimal use
of hardware timers) that---we believe---makes it one of the most
suitable OSes for the embedded and real-time world.

While we weren't able to accurately quantize energy consumption
yet, we can reasonably think that lowering activity of MCU and radio
transceiver will significantly reduce the energy consumption of devices
running RIOT OS. This will be the subject of some of our future
research works.

\bigskip

Currently, RIOT OS supports high-level IoT protocols (6LoWPAN/IPv6, RPL,
TCP, UDP, etc.). However, it still lacks high-performance MAC~/ RDC layer
protocols.

Through this work, we have shown that RIOT OS is also suitable for
implementing high-performance MAC~/ RDC protocols, thanks to its real-time
features (especially hardware timers management).

Moreover, we have improved the robustness of the existing ports of RIOT OS
on MSP430, making it a suitable software platform for tiny motes and devices.




\vfill
\bibliographystyle{apalike}
{\small
\bibliography{sensornets2015}}

\end{document}